\documentclass[twocolumn,showpacs,preprintnumbers,superscriptaddress,prl,floatfix,aps,10pt]{revtex4-1}

\usepackage{hyperref}
\hypersetup{
    bookmarks=true,         
    unicode=true,          
    pdftoolbar=false,        
    pdfmenubar=true,        
    pdffitwindow=true,      
    pdftitle={Anharmonicity changes the solid solubility of a random alloy at high temperatures},    
    pdfauthor={Nina Shulumba},     
    pdfsubject={Subject},   
    pdfnewwindow=true,      
    pdfkeywords={keywords}, 
    colorlinks=true,       
    linkcolor=blue,          
    citecolor=blue,        
    filecolor=black,      
    urlcolor=blue           
    						}
\usepackage{amsmath,amssymb}
\usepackage{graphicx,textcomp}
\usepackage{subfigure}
\usepackage{color}
\usepackage{todonotes}
\usepackage[titletoc,toc,title]{appendix}

\usepackage{xcolor}

\definecolor{olle}{RGB}{240,226,182}
\definecolor{nina}{RGB}{192,225,215}
\definecolor{zamaan}{RGB}{90,150,175}
\definecolor{bjorn}{RGB}{250,160,130}

\usepackage[title,titletoc,toc]{appendix}
\newcommand{\figref}[1]{Fig.~\ref{#1}}

\newcommand{\eqnref}[1]{Eq.~\ref{#1}}

\renewcommand {\vec}    [1]    {\ensuremath{\mathbf{#1}}}

\newcommand   {\mat}    [1]    {\ensuremath{\mathbf{\bar{\bar{#1}}}} }			

\begin{document}

\title{Anharmonicity changes the solid solubility of an alloy at high temperatures} 
\author{Nina Shulumba}
\affiliation{Department of Physics, Chemistry, and Biology (IFM), Link\"{o}ping University, SE-581 83, Link\"oping, Sweden.}
\affiliation{Functional Materials, Saarland University, Campus D3 3, D-66123 Saarbr\"{u}cken, Germany.}

\author{Olle Hellman}
\affiliation{Division of Engineering and Applied Science, California Institute of Technology, Pasadena, California 91125.}
\affiliation{Department of Physics, Chemistry, and Biology (IFM), Link\"{o}ping University, SE-581 83, Link\"oping, Sweden.}

\author{Zamaan Raza}
\affiliation{Department of Physics, Chemistry, and Biology (IFM), Link\"{o}ping University, SE-581 83, Link\"oping, Sweden.}

\author{Bj\"{o}rn Alling}
\affiliation{Department of Physics, Chemistry, and Biology (IFM), Link\"{o}ping University, SE-581 83, Link\"oping, Sweden.}
\affiliation{Max-Planck-Institut f\"{u}r Eisenforschung GmbH, D-40237 D\"{u}sseldorf, Germany}

\author{Jenifer Barrirero}
\affiliation{Functional Materials, Saarland University, Campus D3 3, D-66123 Saarbr\"{u}cken, Germany.}
\affiliation{Department of Physics, Chemistry, and Biology (IFM), Link\"{o}ping University, SE-581 83, Link\"oping, Sweden.}

\author{Frank M\"{u}cklich}
\affiliation{Functional Materials, Saarland University, Campus D3 3, D-66123 Saarbr\"{u}cken, Germany.}

\author{Igor A. Abrikosov}
\affiliation{Department of Physics, Chemistry, and Biology (IFM), Link\"{o}ping University, SE-581 83, Link\"oping, Sweden.}
\affiliation{Materials Modeling and Development Laboratory, NUST ``MISIS'', 119049 Moscow, Russia.}
\affiliation{LOCOMAS Laboratory, Tomsk State University, 634050, Tomsk, Russia}

\author{Magnus Od\'{e}n}
\affiliation{Department of Physics, Chemistry, and Biology (IFM), Link\"{o}ping University, SE-581 83, Link\"oping, Sweden.}

\date{\today}

\begin{abstract}
We have developed a method to accurately and efficiently determine the vibrational free energy as a function of temperature and volume for substitutional alloys from first principles. Taking Ti$_{1-x}$Al$_x$N alloy as a model system, we calculate the isostructural phase diagram by finding the global minimum of the free energy, corresponding to the true equilibrium state of the system. We demonstrate that the anharmonic contribution and temperature dependence of the mixing enthalpy have a decisive impact on the calculated phase diagram of a Ti$_{1-x}$Al$_x$N alloy, lowering the maximum temperature for the miscibility gap from 6560 K to 2860 K. Our local chemical composition measurements on thermally aged Ti$_{0.5}$Al$_{0.5}$N alloys agree with the calculated phase diagram.

\end{abstract}

\maketitle

When discussing the solubility of an alloy, the configurational entropy is always taken into account, but the effect of temperature associated with lattice vibrations is often neglected. It has been experimentally shown \cite{Anthony1993,Fultz1995,Anthony1994} that the vibrational and configurational entropies are comparable in the cases of fcc Ni$_3$Al and Cu$_3$Au, and bcc Fe$_3$Al, and theoretical studies draw the same conclusion \cite{Althoff1997,Ozolins2001,Fultz2010a,Walle2002a,VandeWalle1998} --- that lattice vibrations cannot be ignored. High quality phonon spectra can be computed in the framework of density functional perturbation theory (DFPT) \cite{Baroni2001} and the small displacement method \cite{Alfe2009}, but the introduction of substitutional disorder as in an alloy causes the cost of such calculations to escalate rapidly. In this work, we propose a method of computing the vibrational free energy of a configurationally disordered solid based on the temperature dependent effective potential method (TDEP) \cite{Hellman2011, Hellman2013}, which has an efficiency comparable to the state-of-the-art methods that only apply to ordered solids. Moreover, the method has the advantage of taking into account  anharmonicity of the lattice vibrations and therefore remains valid at temperatures for which the quasi-harmonic approximation breaks down.

We demonstrate the accuracy of our technique in a study of decomposition thermodynamics of Ti$_x$Al$_{1-x}$N alloys \cite{Mayrhofer2003,Abrikosov2011}, a system for which lattice vibrations underpin an unusual and technologically useful isostructural decomposition \cite{Horling2005}. Metastable Ti$_x$Al$_{1-x}$N coatings are ideal for use in the manufacture of cutting tools due to their characteristic age hardening during use. Metastable cubic TiAlN undergoes spinodal decomposition to form nano-scale domains of cubic TiN and AlN, through which extra stress is required to propagate dislocations \cite{Mayrhofer2003,Rachbauer2011,JohanssonJoesaar2013}. Remarkably, calculated values of the maximum temperature for the miscibility gap vary between approximately 6050 and 9000 K \citep{Alling2009,Alling2011}, depending on the methodological details, and as low as 3790 K within the Debye-Gr\"{u}neisen approximation \cite{Wang2012a}. These are well above the dissociation temperatures of TiN and AlN, however, but cutting tools may reach temperatures of up to 1300 K \cite{Norrby2014}, at which point vibrations could be of considerable importance and are subject to the effects of thermal expansion and anharmonicity. We note that existing studies either neglect or use an incomplete description of the vibrational contribution to the free energy, since the methodological challenge and computational efforts required to calculate the phonon spectra of a substitutionally disordered solid using \textit{ab initio} approaches are considerable. 

In this Letter, we propose a computationally tractable method for the treatment of vibrational free energy of a random alloy, and use it to perform accurate first-principles calculations of the vibrational free energy of B1 Ti$_{1-x}$Al$_x$N alloy, our model system. The theoretical miscibility gap has a maximum temperature of 2860 K, and the solubility limit of Al in TiN at intermediate temperatures is increased in comparison to calculations which neglect the effect of lattice vibrations. Our method employs the TDEP method to compute the vibrational contribution to the free energy. When constructing the phase diagram we minimize the Gibbs free energy \cite{Gibbs1871} to obtain the stable alloy compositions at equilibrium. We perform atom probe tomography experiments, exploiting the high spatial and chemical resolutions to depict the decomposition of a supersaturated solid solution of TiAlN with an alloy composition inside the miscibility gap, thus verifying the predicted phase diagram, and demonstrating the importance of lattice vibrations.

Phase stability at constant temperature and volume is determined by the Helmholtz free energy $F$, which can be expressed as the sum of the free energy of a model system $F_m$ and a correction term based on the Kirkwood coupling theorem \cite{Frenkel2002}, where $U_m$ and $U$ are potential energies of model and real system respectively.
\begin{equation}\label{eq:total}
F  =F_m+ \underbrace{  \int_0^1 \left\langle U-U_m \right\rangle_\lambda d\lambda }_{\Delta F}.
\end{equation}
\eqnref{eq:total} is formally exact for phase equilibria in alloys at relevant temperatures. In practice, calculating $\Delta F$ can be very difficult \cite{Grabowski2009}, so our strategy will be to choose the model system with the smallest possible $\Delta F$. Ideally, it should be within the error bars of \emph{ab initio} calculations.

Our model system consists of atoms distributed on the sites of a special quasirandom structure \cite{Zunger1990} (SQS), although any other supercell approximation of the completely random alloy could be used. These atoms, however, interact with effective force constants which have full symmetry of the underlying crystal lattice. Lattice vibrations of the model system are described by temperature dependent effective potential (TDEP) model Hamiltonian \cite{Hellman2011, Hellman2013}:
\begin{equation}\label{eq:harmhamiltonian}
\hat{H}=U_0+\sum_{i} \frac{\vec p_i^2}{2m_i}+\frac{1}{2} \sum_{ij} \vec u_{i} \mat {\Phi}^{\textrm{eff}}_{ij}  \vec u_{j},
\end{equation}
where $\vec{p}_i$ and $\vec{u}_i$ are the momentum and displacement of atom $i$, $\mat {\Phi}^{\textrm{eff}}_{ij}$ are the second order effective force constant tensors, which relate the displacement of atom $j$ to the force $\vec{f}_i$ exerted on atom $i$. The TDEP method allows us to fit parameters of the model Hamiltonian to results of \emph{ab initio} molecular dynamics simulations carried out for the real system of interest. 

In the model Hamiltonian (\eqnref{eq:harmhamiltonian}), the translational and spatial symmetry of the underlying crystal lattice are imposed by treating the components of the alloy as symmetry equivalent, and the real interactions present in DFT are mapped to effective interactions expressed by the interatomic effective alloy force constants $\mat{\Phi}^{\textrm{eff}}_{ij}$ corresponding to the underlying crystal lattice, B1 in case of Ti$_{1-x}$Al$_x$N. Imposing the symmetry reduces the number of independent components in $\mat{\Phi}^{\textrm{eff}}_{ij}$ considerably, making the method numerically efficient \cite{Hellman2013} thus we call our approach the symmetry imposed force constants TDEP method (SIFC-TDEP). In our approach, the forces are obtained from calculations of the real disordered system as described by the SQS and only then mapped onto effective force constants in a system with randomly distributed atomic species having their respective correct atomic masses. Note that the accuracy of the free energy calculations can always be improved, for example, by using a more complex model Hamiltonian. This can be achieved using an expansion of the pairwise force constants AB, BC, AC\ldots resulting in different pairs for different pairs of species, although this is not necessary for our model system.

In order to find the effective force constant matrix that best represents the Born-Oppenheimer potential energy surface, we minimize the difference in forces between the model system and the SQS model of a real alloy, computing the latter by means of \textit{ab initio} molecular dynamics (see supplementary material).

We calculate the internal energy via Density functional theory (DFT), with the electronic contribution to the entropy given by the Mermin functional \cite{Mermin1965}. The configurational entropy is estimated using the mean-field approximation \cite{FurEisenforschung1976}. Short-range clustering effects in Ti$_{1-x}$Al$_x$N have been shown to be significant and similar in magnitude to the mean field contribution at temperatures below 3500 K \cite{Alling2011}. In principle, such clustering effects should be included in the free energy calculations. However, in Ti$_{1-x}$Al$_x$N they were found to predominantly impact the AlN-rich compositions, having little effect on the solubility limits in the TiN-rich region, and only moderately decrease the maximum temperature of the miscibility gap. For these reasons, short-range clustering are not included in our calculations.

\begin{figure}
\includegraphics[width=\linewidth]{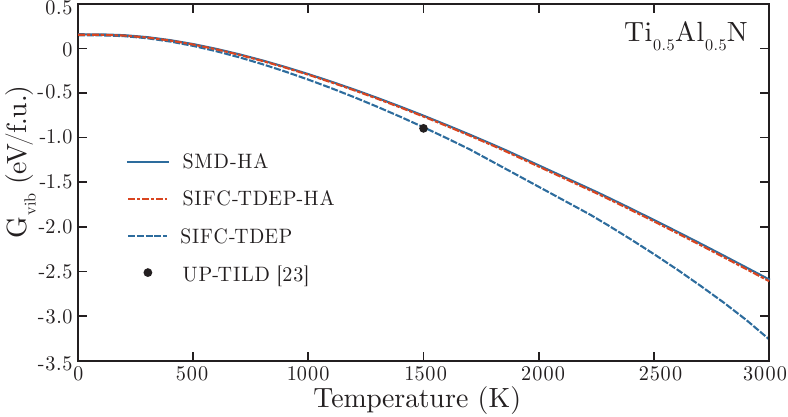}
\caption{\label{fig:vibs}  (Color online) Comparison of harmonic and anharmonic vibrational Gibbs free energies as a function of temperature. Harmonic free energies were calculated via the TDEP (SIFC-TDEP-HA) and small displacement (SMD-HA) methods.  The free energy including anharmonic effects via the terms $\Delta U_0$ and $g(\omega,T)$ (SIFC-TDEP) is compared with a single thermodynamic integration data point.}
\end{figure}

We decompose the Helmholtz free energy into vibrational (``vib''), configurational (``config'') and electronic terms (``el''):
\begin{equation}\label{totalF}
F  =U-TS=F_{\mathrm{el}}+F_{\mathrm{vib}}+F_{\mathrm{config}}
\end{equation}
Using the force constants in \eqnref{eq:harmhamiltonian}, we can calculate the phonon density of states $g(\omega,T)$, from which we compute the vibrational contribution to the Helmholtz free energy:
\begin{equation}\label{eq:phonon_free_energy}
\begin{split}
& F_{\textrm{vib}}(\omega,T)=\Delta U_0(T)+\int_0^\infty g(\omega,T)\frac{\hbar \omega}{2} d\omega+\\
& +\int_0^\infty g(\omega,T)k_\textrm{B} T \ln \left( 1- \exp \left( -\frac{\hbar\omega}{k_\textrm{B}T} \right) \right) d\omega.
 \end{split}
\end{equation}
In our case $g(\omega,T)$ is strongly temperature dependent and includes anharmonic effects and thermal expansion. $\Delta U_0$ is an anharmonic correction to the energy in TDEP formalism that is added to F$_{vib}$ (see supplementary material). 

The first test of our model system was to see that at low temperatures it does not deviate from a direct harmonic calculation using standard methods \cite{Baroni2001,Alfe2009}, as illustrated in \figref{fig:vibs}. The vibrational free energy of our model system agrees well with the harmonic model, with a difference of 2 meV/atom at 0 K and 4 meV/atom at 1500 K. For this to be a fair test, we used the TDEP force constants from 300 K where anharmonic contributions are expected to be negligible. This was a probe of the error introduced by the imposed symmetry, which turned out to be small.

At higher temperatures, the harmonic model is no longer a good benchmark, and we calculate the formally exact free energy difference using thermodynamic integration (UP-TILD)~\cite{Grabowski2009}. Due to its considerable computational cost, UP-TILD was applied to a single point, for which anharmonic effects are considerable. The correction to the free energy (\eqnref{eq:total}) at 1500 K was calculated to be $\sim$~5~meV/atom. We note that the harmonic approximation is $\sim$~15~meV/atom further from the formally exact number (See Fig. 1 in supplementary material). We conclude that the SIFC-TDEP and UP-TILD free energies are within our desired error bars, and can use our model system free energies for the subsequent calculations with confidence.

We construct the Gibbs free energy ($G=F+PV$) from the Helmholtz free energy surface, where $P=-dF/dV$. The Gibbs free energy of mixing is given by,
\begin{equation}\label{eq:gmix}
G_{\textrm{mix}}(T,x)=G_{\textrm{Ti}_{1-x}\textrm{Al}_x\textrm{N}}-(1-x)G_{\textrm{TiN}}-xG_{\textrm{AlN}},
\end{equation}
where $x$ is the fraction of AlN.

\begin{figure}
\includegraphics[width=\linewidth]{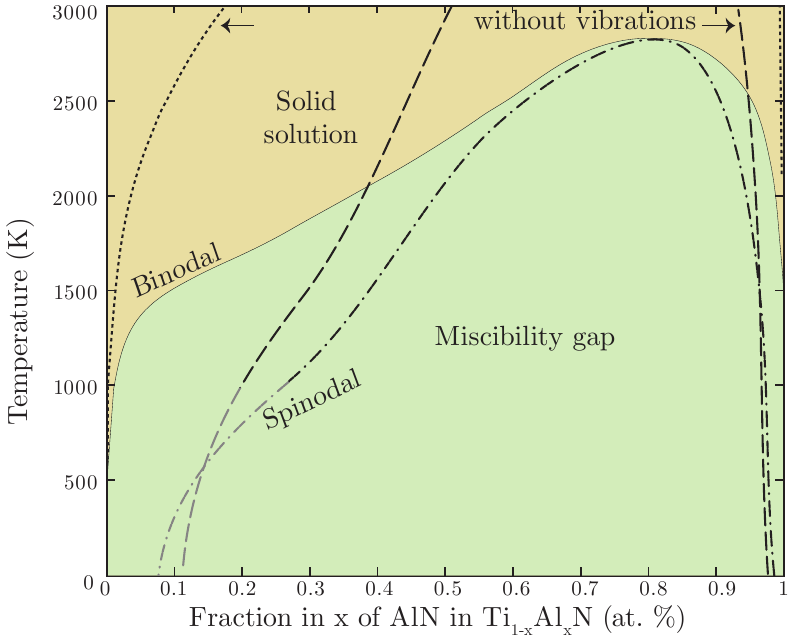}
\caption{\label{fig:PD}  (Color online) Calculated phase diagram for B1-Ti$_{1-x}$Al$_{x}$N. Comparing the dotted and solid lines shows the effect of including the vibrational entropy for binodal using the SIFC-TDEP method. The dash-dotted line corresponds to the spinodal metastable region including the anharmonic contribution, and the dashed line in solid solution region to the spinodal line without the vibrational contribution.} 
\end{figure}

We reconstruct the concentration-temperature phase diagram at zero pressure by a direct minimization of $G_\mathrm{mix}$ at each temperature and global composition on a grid (see supplementary material). We determine the spinodal from the condition ($\frac{\partial^2 G_{mix}}{\partial x^2}|_{T} \leq 0$). 

The calculated phase diagram is compared with one neglecting lattice vibrations in \figref{fig:PD}. The difference in both the spinodal region and the miscibility gap is dramatic. By including the vibrational contribution to the free energy, the maximum of the miscibility gap is lowered from 6560 K to 2860 K. The roots of the second derivative of the total Gibbs free energy below 1000 K depend on the fitting function, in this case third order Redlich-Kister polynomials \cite{Redlich1948} (gray lines of spinodal).
t
\begin{figure}
\includegraphics[width=\linewidth]{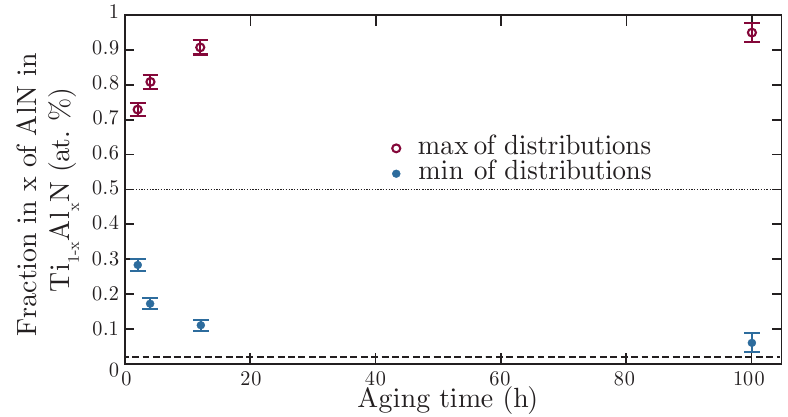}
\caption{\label{fig:normdist}  (Color online) The limits of the concentration of metal sublattices as a function of aging time.} 
\end{figure}

Experimental verification of theoretical phase diagram (\figref{fig:PD}) is a challenging task. Because of the high melting temperature of TiN, its alloys are synthesized by thin-film deposition techniques, often followed by an annealing.  Due to the explicit out-of-equilibrium nature of this metastable transformation, direct information from equilibrium experiments is not available, so we adopted the following strategy. 4 $\mu$m thick TiAlN films with equal amounts of Ti and Al ($x=0.5$) were deposited by cathodic arc evaporation, resulting in a nearly uniform solid solution \cite{Johnson2012}. Our theoretical predictions suggest that when diffusion is activated this alloy will phase segregate through spinodal decomposition. To study this effect, a set of samples was prepared by annealing the TiAlN-films at 1073 K for 2, 4, 12 and 100 hours in an argon atmosphere. The evolved microstructures where then examined by atom probe tomography.

\figref{fig:2Dplots} shows 2D compositional maps of 2 nm thick slices through the reconstructed 3D atom probe specimen of the annealed samples. The initially homogeneous cubic solid solution decomposes isostructurally forming cubic Ti- and Al-rich domains during  the first 12 hours. After 100 hours, the metastable cubic AlN-rich phase has to a large extent transformed to its stable wurtzite phase \cite{Norrby2014}. During the first 12 hours the evolution of the microstructure and the chemical segregation becomes gradually more pronounced, which is consistent with the spinodal decomposition \cite{Knutsson2013}. After 100 hours the microstructure is coarser.

The phase diagram (\figref{fig:PD}) shows a substantial shift in the solubility at 1073 K compared to the predictions excluding vibrational effects, especially for the solubility of Al in TiN. In order to experimentally verify this, we extracted the maximum and minimum local composition in the decomposed microstructure by constructing histograms of the concentrations in 1 nm$^3$ bins throughout the sample. These two local compositional extrema are plotted in \figref{fig:normdist} as a function of annealing time, and demonstrate the timescale and decomposition path of the alloy. The solubility of Al in TiN at 1073 K tends to a final solubility of $\sim$2 \% according to the theoretical phase diagram. The solubility limit for Ti in cubic AlN could not be determined experimentally since it was not reached prior to the onset of wurtzite AlN formation. However, the experimental data for Ti in cubic AlN asymptotically approaches the theoretical value ($x=1$). 

In summary, we present an accurate technique for calculating the vibrational contribution to the Gibbs free energy for random alloys. We reconstruct the phase diagram of a model Ti$_{1-x}$Al$_x$N alloy, and demonstrate that in this system, the non-harmonic vibrational phonon free energy is large and comparable to the harmonic energies. As a result, we find a dramatic decrease of the maximum temperature for the miscibility gap, from 6560 K to 2860 K, as well as an increase AlN of solubility in TiN as compared to calculations which neglect lattice vibrations. Atom probe tomography experiments on annealed Ti$_{0.5}$Al$_{0.5}$N samples are in line with our theoretical predictions demonstrating a finite AlN content in the TiN rich compositions after 100 h of annealing at 1073 K.

\textsc{Acknowledgments}:
This work was supported by the Swedish Foundation for Strategic Research (SSF) programs RMA08---0069, SRL10---0026 and FUNCASE, the Swedish Research Council (VR) projects 2012---4401, 621-2011-4426, the Swedish Governmental Agency for Innovation Systems (Vinnova) through the M---ERA.net project MC$_2$ and SECO Tools AB. Support from the Swedish Research Council (VR) program 637-2013-7296 is gratefully acknowledged by O. Hellman. B. Alling acknowledges financial support from the Swedish Research Council (VR) through grant 621-2011-4417 and 330-2014-6336. N. Shulumba and J. Barrirero acknowledge the financial support from the Erasmus Mundus Joint European Doctoral Programme DocMASE. I. A. Abrikosov acknowledges the support from the Grant of Ministry of Education and Science of the Russian Federation (Grant No. 14.Y26.31.0005) and Tomsk State University Academic D. I. Mendeleev Fund Program. We thank Isabella Schramm (Saarland University) for the assistance with atom probe tomography measurements. All calculations were performed using the supercomputer resources of the Swedish National Infrastructure for Computing (SNIC), PDC and NSC centers.


\begin{thebibliography}{30}%
\makeatletter
\providecommand \@ifxundefined [1]{%
 \@ifx{#1\undefined}
}%
\providecommand \@ifnum [1]{%
 \ifnum #1\expandafter \@firstoftwo
 \else \expandafter \@secondoftwo
 \fi
}%
\providecommand \@ifx [1]{%
 \ifx #1\expandafter \@firstoftwo
 \else \expandafter \@secondoftwo
 \fi
}%
\providecommand \natexlab [1]{#1}%
\providecommand \enquote  [1]{``#1''}%
\providecommand \bibnamefont  [1]{#1}%
\providecommand \bibfnamefont [1]{#1}%
\providecommand \citenamefont [1]{#1}%
\providecommand \href@noop [0]{\@secondoftwo}%
\providecommand \href [0]{\begingroup \@sanitize@url \@href}%
\providecommand \@href[1]{\@@startlink{#1}\@@href}%
\providecommand \@@href[1]{\endgroup#1\@@endlink}%
\providecommand \@sanitize@url [0]{\catcode `\\12\catcode `\$12\catcode
  `\&12\catcode `\#12\catcode `\^12\catcode `\_12\catcode `\%12\relax}%
\providecommand \@@startlink[1]{}%
\providecommand \@@endlink[0]{}%
\providecommand \url  [0]{\begingroup\@sanitize@url \@url }%
\providecommand \@url [1]{\endgroup\@href {#1}{\urlprefix }}%
\providecommand \urlprefix  [0]{URL }%
\providecommand \Eprint [0]{\href }%
\providecommand \doibase [0]{http://dx.doi.org/}%
\providecommand \selectlanguage [0]{\@gobble}%
\providecommand \bibinfo  [0]{\@secondoftwo}%
\providecommand \bibfield  [0]{\@secondoftwo}%
\providecommand \translation [1]{[#1]}%
\providecommand \BibitemOpen [0]{}%
\providecommand \bibitemStop [0]{}%
\providecommand \bibitemNoStop [0]{.\EOS\space}%
\providecommand \EOS [0]{\spacefactor3000\relax}%
\providecommand \BibitemShut  [1]{\csname bibitem#1\endcsname}%
\let\auto@bib@innerbib\@empty
\bibitem [{\citenamefont {Anthony}\ \emph {et~al.}(1993)\citenamefont
  {Anthony}, \citenamefont {Okamoto},\ and\ \citenamefont
  {Fultz}}]{Anthony1993}%
  \BibitemOpen
  \bibfield  {author} {\bibinfo {author} {\bibfnamefont {L.}~\bibnamefont
  {Anthony}}, \bibinfo {author} {\bibfnamefont {J.}~\bibnamefont {Okamoto}}, \
  and\ \bibinfo {author} {\bibfnamefont {B.}~\bibnamefont {Fultz}},\ }\href
  {\doibase 10.1103/PhysRevLett.70.1128} {\bibfield  {journal} {\bibinfo
  {journal} {Physical Review Letters}\ }\textbf {\bibinfo {volume} {70}},\
  \bibinfo {pages} {1128} (\bibinfo {year} {1993})}\BibitemShut {NoStop}%
\bibitem [{\citenamefont {Fultz}\ \emph {et~al.}(1995)\citenamefont {Fultz},
  \citenamefont {Anthony}, \citenamefont {Nagel}, \citenamefont {Nicklow},\
  and\ \citenamefont {Spooner}}]{Fultz1995}%
  \BibitemOpen
  \bibfield  {author} {\bibinfo {author} {\bibfnamefont {B.}~\bibnamefont
  {Fultz}}, \bibinfo {author} {\bibfnamefont {L.}~\bibnamefont {Anthony}},
  \bibinfo {author} {\bibfnamefont {L.~J.}\ \bibnamefont {Nagel}}, \bibinfo
  {author} {\bibfnamefont {R.~M.}\ \bibnamefont {Nicklow}}, \ and\ \bibinfo
  {author} {\bibfnamefont {S.}~\bibnamefont {Spooner}},\ }\href {\doibase
  10.1103/PhysRevB.52.3315} {\bibfield  {journal} {\bibinfo  {journal}
  {Physical Review B}\ }\textbf {\bibinfo {volume} {52}},\ \bibinfo {pages}
  {3315} (\bibinfo {year} {1995})}\BibitemShut {NoStop}%
\bibitem [{\citenamefont {Anthony}\ \emph {et~al.}(1994)\citenamefont
  {Anthony}, \citenamefont {Nagel}, \citenamefont {Okamoto},\ and\
  \citenamefont {Fultz}}]{Anthony1994}%
  \BibitemOpen
  \bibfield  {author} {\bibinfo {author} {\bibfnamefont {L.}~\bibnamefont
  {Anthony}}, \bibinfo {author} {\bibfnamefont {L.~J.}\ \bibnamefont {Nagel}},
  \bibinfo {author} {\bibfnamefont {J.~K.}\ \bibnamefont {Okamoto}}, \ and\
  \bibinfo {author} {\bibfnamefont {B.}~\bibnamefont {Fultz}},\ }\href
  {\doibase 10.1103/PhysRevLett.73.3034} {\bibfield  {journal} {\bibinfo
  {journal} {Physical Review Letters}\ }\textbf {\bibinfo {volume} {73}},\
  \bibinfo {pages} {3034} (\bibinfo {year} {1994})}\BibitemShut {NoStop}%
\bibitem [{\citenamefont {Althoff}\ \emph {et~al.}(1997)\citenamefont
  {Althoff}, \citenamefont {Morgan}, \citenamefont {de~Fontaine}, \citenamefont
  {Asta}, \citenamefont {Foiles},\ and\ \citenamefont {Johnson}}]{Althoff1997}%
  \BibitemOpen
  \bibfield  {author} {\bibinfo {author} {\bibfnamefont {J.~D.}\ \bibnamefont
  {Althoff}}, \bibinfo {author} {\bibfnamefont {D.}~\bibnamefont {Morgan}},
  \bibinfo {author} {\bibfnamefont {D.}~\bibnamefont {de~Fontaine}}, \bibinfo
  {author} {\bibfnamefont {M.}~\bibnamefont {Asta}}, \bibinfo {author}
  {\bibfnamefont {S.~M.}\ \bibnamefont {Foiles}}, \ and\ \bibinfo {author}
  {\bibfnamefont {D.~D.}\ \bibnamefont {Johnson}},\ }\href {\doibase
  10.1103/PhysRevB.56.R5705} {\bibfield  {journal} {\bibinfo  {journal}
  {Physical Review B}\ }\textbf {\bibinfo {volume} {56}},\ \bibinfo {pages}
  {R5705} (\bibinfo {year} {1997})}\BibitemShut {NoStop}%
\bibitem [{\citenamefont {Ozoliņs}\ and\ \citenamefont
  {Asta}(2001)}]{Ozolins2001}%
  \BibitemOpen
  \bibfield  {author} {\bibinfo {author} {\bibfnamefont {V.}~\bibnamefont
  {Ozoliņs}}\ and\ \bibinfo {author} {\bibfnamefont {M.}~\bibnamefont
  {Asta}},\ }\href {\doibase 10.1103/PhysRevLett.86.448} {\bibfield  {journal}
  {\bibinfo  {journal} {Physical Review Letters}\ }\textbf {\bibinfo {volume}
  {86}},\ \bibinfo {pages} {448} (\bibinfo {year} {2001})}\BibitemShut
  {NoStop}%
\bibitem [{\citenamefont {Fultz}(2010)}]{Fultz2010a}%
  \BibitemOpen
  \bibfield  {author} {\bibinfo {author} {\bibfnamefont {B.}~\bibnamefont
  {Fultz}},\ }\href {\doibase 10.1016/j.pmatsci.2009.05.002} {\bibfield
  {journal} {\bibinfo  {journal} {Progress in Materials Science}\ }\textbf
  {\bibinfo {volume} {55}},\ \bibinfo {pages} {247} (\bibinfo {year}
  {2010})}\BibitemShut {NoStop}%
\bibitem [{\citenamefont {van~de Walle}\ and\ \citenamefont
  {Ceder}(2002)}]{Walle2002a}%
  \BibitemOpen
  \bibfield  {author} {\bibinfo {author} {\bibfnamefont {A.}~\bibnamefont
  {van~de Walle}}\ and\ \bibinfo {author} {\bibfnamefont {G.}~\bibnamefont
  {Ceder}},\ }\href {\doibase 10.1103/RevModPhys.74.11} {\bibfield  {journal}
  {\bibinfo  {journal} {Reviews of Modern Physics}\ }\textbf {\bibinfo {volume}
  {74}},\ \bibinfo {pages} {11} (\bibinfo {year} {2002})}\BibitemShut {NoStop}%
\bibitem [{\citenamefont {van~de Walle}\ \emph {et~al.}(1998)\citenamefont
  {van~de Walle}, \citenamefont {Ceder},\ and\ \citenamefont
  {Waghmare}}]{VandeWalle1998}%
  \BibitemOpen
  \bibfield  {author} {\bibinfo {author} {\bibfnamefont {A.}~\bibnamefont
  {van~de Walle}}, \bibinfo {author} {\bibfnamefont {G.}~\bibnamefont {Ceder}},
  \ and\ \bibinfo {author} {\bibfnamefont {U.~V.}\ \bibnamefont {Waghmare}},\
  }\href {\doibase 10.1103/PhysRevLett.80.4911} {\bibfield  {journal} {\bibinfo
   {journal} {Physical Review Letters}\ }\textbf {\bibinfo {volume} {80}},\
  \bibinfo {pages} {4911} (\bibinfo {year} {1998})}\BibitemShut {NoStop}%
\bibitem [{\citenamefont {Baroni}\ \emph {et~al.}(2001)\citenamefont {Baroni},
  \citenamefont {de~Gironcoli},\ and\ \citenamefont {{Dal
  Corso}}}]{Baroni2001}%
  \BibitemOpen
  \bibfield  {author} {\bibinfo {author} {\bibfnamefont {S.}~\bibnamefont
  {Baroni}}, \bibinfo {author} {\bibfnamefont {S.}~\bibnamefont
  {de~Gironcoli}}, \ and\ \bibinfo {author} {\bibfnamefont {A.}~\bibnamefont
  {{Dal Corso}}},\ }\href {\doibase 10.1103/RevModPhys.73.515} {\bibfield
  {journal} {\bibinfo  {journal} {Reviews of Modern Physics}\ }\textbf
  {\bibinfo {volume} {73}},\ \bibinfo {pages} {515} (\bibinfo {year}
  {2001})}\BibitemShut {NoStop}%
\bibitem [{\citenamefont {Alf\`{e}}(2009)}]{Alfe2009}%
  \BibitemOpen
  \bibfield  {author} {\bibinfo {author} {\bibfnamefont {D.}~\bibnamefont
  {Alf\`{e}}},\ }\href {\doibase 10.1016/j.cpc.2009.03.010} {\bibfield
  {journal} {\bibinfo  {journal} {Computer Physics Communications}\ }\textbf
  {\bibinfo {volume} {180}},\ \bibinfo {pages} {2622} (\bibinfo {year}
  {2009})}\BibitemShut {NoStop}%
\bibitem [{\citenamefont {Hellman}\ \emph {et~al.}(2011)\citenamefont
  {Hellman}, \citenamefont {Abrikosov},\ and\ \citenamefont
  {Simak}}]{Hellman2011}%
  \BibitemOpen
  \bibfield  {author} {\bibinfo {author} {\bibfnamefont {O.}~\bibnamefont
  {Hellman}}, \bibinfo {author} {\bibfnamefont {I.~A.}\ \bibnamefont
  {Abrikosov}}, \ and\ \bibinfo {author} {\bibfnamefont {S.~I.}\ \bibnamefont
  {Simak}},\ }\href {\doibase 10.1103/PhysRevB.84.180301} {\bibfield  {journal}
  {\bibinfo  {journal} {Physical Review B}\ }\textbf {\bibinfo {volume} {84}},\
  \bibinfo {pages} {180301} (\bibinfo {year} {2011})}\BibitemShut {NoStop}%
\bibitem [{\citenamefont {Hellman}\ \emph {et~al.}(2013)\citenamefont
  {Hellman}, \citenamefont {Steneteg}, \citenamefont {Abrikosov},\ and\
  \citenamefont {Simak}}]{Hellman2013}%
  \BibitemOpen
  \bibfield  {author} {\bibinfo {author} {\bibfnamefont {O.}~\bibnamefont
  {Hellman}}, \bibinfo {author} {\bibfnamefont {P.}~\bibnamefont {Steneteg}},
  \bibinfo {author} {\bibfnamefont {I.~A.}\ \bibnamefont {Abrikosov}}, \ and\
  \bibinfo {author} {\bibfnamefont {S.~I.}\ \bibnamefont {Simak}},\ }\href
  {\doibase 10.1103/PhysRevB.87.104111} {\bibfield  {journal} {\bibinfo
  {journal} {Physical Review B}\ }\textbf {\bibinfo {volume} {87}},\ \bibinfo
  {pages} {104111} (\bibinfo {year} {2013})}\BibitemShut {NoStop}%
\bibitem [{\citenamefont {Mayrhofer}\ \emph {et~al.}(2003)\citenamefont
  {Mayrhofer}, \citenamefont {Hörling}, \citenamefont {Karlsson},
  \citenamefont {Sjölén}, \citenamefont {Larsson}, \citenamefont
  {Mitterer},\ and\ \citenamefont {Hultman}}]{Mayrhofer2003}%
  \BibitemOpen
  \bibfield  {author} {\bibinfo {author} {\bibfnamefont {P.~H.}\ \bibnamefont
  {Mayrhofer}}, \bibinfo {author} {\bibfnamefont {A.}~\bibnamefont
  {Hörling}}, \bibinfo {author} {\bibfnamefont {L.}~\bibnamefont {Karlsson}},
  \bibinfo {author} {\bibfnamefont {J.}~\bibnamefont {Sjölén}}, \bibinfo
  {author} {\bibfnamefont {T.}~\bibnamefont {Larsson}}, \bibinfo {author}
  {\bibfnamefont {C.}~\bibnamefont {Mitterer}}, \ and\ \bibinfo {author}
  {\bibfnamefont {L.}~\bibnamefont {Hultman}},\ }\href {\doibase
  10.1063/1.1608464} {\bibfield  {journal} {\bibinfo  {journal} {Applied
  Physics Letters}\ }\textbf {\bibinfo {volume} {83}},\ \bibinfo {pages} {2049}
  (\bibinfo {year} {2003})}\BibitemShut {NoStop}%
\bibitem [{\citenamefont {Abrikosov}\ \emph {et~al.}(2011)\citenamefont
  {Abrikosov}, \citenamefont {Knutsson}, \citenamefont {Alling}, \citenamefont
  {Tasn\'{a}di}, \citenamefont {Lind}, \citenamefont {Hultman},\ and\
  \citenamefont {Od\'{e}n}}]{Abrikosov2011}%
  \BibitemOpen
  \bibfield  {author} {\bibinfo {author} {\bibfnamefont {I.~a.}\ \bibnamefont
  {Abrikosov}}, \bibinfo {author} {\bibfnamefont {A.}~\bibnamefont {Knutsson}},
  \bibinfo {author} {\bibfnamefont {B.}~\bibnamefont {Alling}}, \bibinfo
  {author} {\bibfnamefont {F.}~\bibnamefont {Tasn\'{a}di}}, \bibinfo {author}
  {\bibfnamefont {H.}~\bibnamefont {Lind}}, \bibinfo {author} {\bibfnamefont
  {L.}~\bibnamefont {Hultman}}, \ and\ \bibinfo {author} {\bibfnamefont
  {M.}~\bibnamefont {Od\'{e}n}},\ }\href {\doibase 10.3390/ma4091599}
  {\bibfield  {journal} {\bibinfo  {journal} {Materials}\ }\textbf {\bibinfo
  {volume} {4}},\ \bibinfo {pages} {1599} (\bibinfo {year} {2011})}\BibitemShut
  {NoStop}%
\bibitem [{\citenamefont {H\"{o}rling}\ \emph {et~al.}(2005)\citenamefont
  {H\"{o}rling}, \citenamefont {Hultman}, \citenamefont {Od\'{e}n},
  \citenamefont {Sj\"{o}l\'{e}n},\ and\ \citenamefont
  {Karlsson}}]{Horling2005}%
  \BibitemOpen
  \bibfield  {author} {\bibinfo {author} {\bibfnamefont {A.}~\bibnamefont
  {H\"{o}rling}}, \bibinfo {author} {\bibfnamefont {L.}~\bibnamefont
  {Hultman}}, \bibinfo {author} {\bibfnamefont {M.}~\bibnamefont {Od\'{e}n}},
  \bibinfo {author} {\bibfnamefont {J.}~\bibnamefont {Sj\"{o}l\'{e}n}}, \ and\
  \bibinfo {author} {\bibfnamefont {L.}~\bibnamefont {Karlsson}},\ }\href
  {\doibase 10.1016/j.surfcoat.2004.04.056} {\bibfield  {journal} {\bibinfo
  {journal} {Surface and Coatings Technology}\ }\textbf {\bibinfo {volume}
  {191}},\ \bibinfo {pages} {384} (\bibinfo {year} {2005})}\BibitemShut
  {NoStop}%
\bibitem [{\citenamefont {Rachbauer}\ \emph {et~al.}(2011)\citenamefont
  {Rachbauer}, \citenamefont {Massl}, \citenamefont {Stergar}, \citenamefont
  {Holec}, \citenamefont {Kiener}, \citenamefont {Keckes}, \citenamefont
  {Patscheider}, \citenamefont {Stiefel}, \citenamefont {Leitner},\ and\
  \citenamefont {Mayrhofer}}]{Rachbauer2011}%
  \BibitemOpen
  \bibfield  {author} {\bibinfo {author} {\bibfnamefont {R.}~\bibnamefont
  {Rachbauer}}, \bibinfo {author} {\bibfnamefont {S.}~\bibnamefont {Massl}},
  \bibinfo {author} {\bibfnamefont {E.}~\bibnamefont {Stergar}}, \bibinfo
  {author} {\bibfnamefont {D.}~\bibnamefont {Holec}}, \bibinfo {author}
  {\bibfnamefont {D.}~\bibnamefont {Kiener}}, \bibinfo {author} {\bibfnamefont
  {J.}~\bibnamefont {Keckes}}, \bibinfo {author} {\bibfnamefont
  {J.}~\bibnamefont {Patscheider}}, \bibinfo {author} {\bibfnamefont
  {M.}~\bibnamefont {Stiefel}}, \bibinfo {author} {\bibfnamefont
  {H.}~\bibnamefont {Leitner}}, \ and\ \bibinfo {author} {\bibfnamefont
  {P.~H.}\ \bibnamefont {Mayrhofer}},\ }\href {\doibase 10.1063/1.3610451}
  {\bibfield  {journal} {\bibinfo  {journal} {Journal of Applied Physics}\
  }\textbf {\bibinfo {volume} {110}},\ \bibinfo {pages} {023515} (\bibinfo
  {year} {2011})}\BibitemShut {NoStop}%
\bibitem [{\citenamefont {{Johansson J\~{o}esaar}}\ \emph
  {et~al.}(2013)\citenamefont {{Johansson J\~{o}esaar}}, \citenamefont
  {Norrby}, \citenamefont {Ullbrand}, \citenamefont {M'Saoubi},\ and\
  \citenamefont {Od\'{e}n}}]{JohanssonJoesaar2013}%
  \BibitemOpen
  \bibfield  {author} {\bibinfo {author} {\bibfnamefont {M.~P.}\ \bibnamefont
  {{Johansson J\~{o}esaar}}}, \bibinfo {author} {\bibfnamefont
  {N.}~\bibnamefont {Norrby}}, \bibinfo {author} {\bibfnamefont
  {J.}~\bibnamefont {Ullbrand}}, \bibinfo {author} {\bibfnamefont
  {R.}~\bibnamefont {M'Saoubi}}, \ and\ \bibinfo {author} {\bibfnamefont
  {M.}~\bibnamefont {Od\'{e}n}},\ }\href {\doibase
  10.1016/j.surfcoat.2013.07.031} {\bibfield  {journal} {\bibinfo  {journal}
  {Surface and Coatings Technology}\ }\textbf {\bibinfo {volume} {235}},\
  \bibinfo {pages} {181} (\bibinfo {year} {2013})}\BibitemShut {NoStop}%
\bibitem [{\citenamefont {Alling}\ \emph {et~al.}(2009)\citenamefont {Alling},
  \citenamefont {Odén}, \citenamefont {Hultman},\ and\ \citenamefont
  {Abrikosov}}]{Alling2009}%
  \BibitemOpen
  \bibfield  {author} {\bibinfo {author} {\bibfnamefont {B.}~\bibnamefont
  {Alling}}, \bibinfo {author} {\bibfnamefont {M.}~\bibnamefont {Odén}},
  \bibinfo {author} {\bibfnamefont {L.}~\bibnamefont {Hultman}}, \ and\
  \bibinfo {author} {\bibfnamefont {I.~A.}\ \bibnamefont {Abrikosov}},\ }\href
  {\doibase 10.1063/1.3256196} {\bibfield  {journal} {\bibinfo  {journal}
  {Applied Physics Letters}\ }\textbf {\bibinfo {volume} {95}},\ \bibinfo
  {pages} {181906} (\bibinfo {year} {2009})}\BibitemShut {NoStop}%
\bibitem [{\citenamefont {Alling}\ \emph {et~al.}(2011)\citenamefont {Alling},
  \citenamefont {Ruban}, \citenamefont {Karimi}, \citenamefont {Hultman},\ and\
  \citenamefont {Abrikosov}}]{Alling2011}%
  \BibitemOpen
  \bibfield  {author} {\bibinfo {author} {\bibfnamefont {B.}~\bibnamefont
  {Alling}}, \bibinfo {author} {\bibfnamefont {A.~V.}\ \bibnamefont {Ruban}},
  \bibinfo {author} {\bibfnamefont {A.}~\bibnamefont {Karimi}}, \bibinfo
  {author} {\bibfnamefont {L.}~\bibnamefont {Hultman}}, \ and\ \bibinfo
  {author} {\bibfnamefont {I.~A.}\ \bibnamefont {Abrikosov}},\ }\href {\doibase
  10.1103/PhysRevB.83.104203} {\bibfield  {journal} {\bibinfo  {journal}
  {Physical Review B}\ }\textbf {\bibinfo {volume} {83}},\ \bibinfo {pages}
  {104203} (\bibinfo {year} {2011})}\BibitemShut {NoStop}%
\bibitem [{\citenamefont {Wang}\ \emph {et~al.}(2012)\citenamefont {Wang},
  \citenamefont {Shang}, \citenamefont {Du}, \citenamefont {Chen},
  \citenamefont {Wang},\ and\ \citenamefont {Liu}}]{Wang2012a}%
  \BibitemOpen
  \bibfield  {author} {\bibinfo {author} {\bibfnamefont {A.}~\bibnamefont
  {Wang}}, \bibinfo {author} {\bibfnamefont {S.-L.}\ \bibnamefont {Shang}},
  \bibinfo {author} {\bibfnamefont {Y.}~\bibnamefont {Du}}, \bibinfo {author}
  {\bibfnamefont {L.}~\bibnamefont {Chen}}, \bibinfo {author} {\bibfnamefont
  {J.}~\bibnamefont {Wang}}, \ and\ \bibinfo {author} {\bibfnamefont {Z.-K.}\
  \bibnamefont {Liu}},\ }\href {\doibase 10.1007/s10853-011-6223-z} {\bibfield
  {journal} {\bibinfo  {journal} {Journal of Materials Science}\ }\textbf
  {\bibinfo {volume} {47}},\ \bibinfo {pages} {7621} (\bibinfo {year}
  {2012})}\BibitemShut {NoStop}%
\bibitem [{\citenamefont {Norrby}\ \emph {et~al.}(2014)\citenamefont {Norrby},
  \citenamefont {Rogstr\"{o}m}, \citenamefont {Johansson-J\~{o}esaar},
  \citenamefont {Schell},\ and\ \citenamefont {Od\'{e}n}}]{Norrby2014}%
  \BibitemOpen
  \bibfield  {author} {\bibinfo {author} {\bibfnamefont {N.}~\bibnamefont
  {Norrby}}, \bibinfo {author} {\bibfnamefont {L.}~\bibnamefont
  {Rogstr\"{o}m}}, \bibinfo {author} {\bibfnamefont {M.}~\bibnamefont
  {Johansson-J\~{o}esaar}}, \bibinfo {author} {\bibfnamefont {N.}~\bibnamefont
  {Schell}}, \ and\ \bibinfo {author} {\bibfnamefont {M.}~\bibnamefont
  {Od\'{e}n}},\ }\href {\doibase 10.1016/j.actamat.2014.04.014} {\bibfield
  {journal} {\bibinfo  {journal} {Acta Materialia}\ }\textbf {\bibinfo {volume}
  {73}},\ \bibinfo {pages} {205} (\bibinfo {year} {2014})}\BibitemShut
  {NoStop}%
\bibitem [{\citenamefont {Gibbs}(1871)}]{Gibbs1871}%
  \BibitemOpen
  \bibfield  {author} {\bibinfo {author} {\bibfnamefont {J.~W.}\ \bibnamefont
  {Gibbs}},\ }\href {http://books.google.se/books?id=6ijzXwAACAAJ} {\emph
  {\bibinfo {title} {{A Method of Geometrical Representation of the
  Thermodynamic Properties of Substances by Means of Surfaces}}}},\ Vol.\
  \bibinfo {volume} {Volumes 38}\ (\bibinfo  {publisher} {The Academy},\
  \bibinfo {year} {1871})\BibitemShut {NoStop}%
\bibitem [{\citenamefont {Frenkel}\ \emph {et~al.}(1997)\citenamefont
  {Frenkel}, \citenamefont {Smit},\ and\ \citenamefont {Ratner}}]{Frenkel2002}%
  \BibitemOpen
  \bibfield  {author} {\bibinfo {author} {\bibfnamefont {D.}~\bibnamefont
  {Frenkel}}, \bibinfo {author} {\bibfnamefont {B.}~\bibnamefont {Smit}}, \
  and\ \bibinfo {author} {\bibfnamefont {M.~A.}\ \bibnamefont {Ratner}},\
  }\href {\doibase 10.1063/1.881812} {\bibfield  {journal} {\bibinfo  {journal}
  {Physics Today}\ }\textbf {\bibinfo {volume} {50}},\ \bibinfo {pages} {66}
  (\bibinfo {year} {1997})}\BibitemShut {NoStop}%
\bibitem [{\citenamefont {Grabowski}\ \emph {et~al.}(2009)\citenamefont
  {Grabowski}, \citenamefont {Ismer}, \citenamefont {Hickel},\ and\
  \citenamefont {Neugebauer}}]{Grabowski2009}%
  \BibitemOpen
  \bibfield  {author} {\bibinfo {author} {\bibfnamefont {B.}~\bibnamefont
  {Grabowski}}, \bibinfo {author} {\bibfnamefont {L.}~\bibnamefont {Ismer}},
  \bibinfo {author} {\bibfnamefont {T.}~\bibnamefont {Hickel}}, \ and\ \bibinfo
  {author} {\bibfnamefont {J.}~\bibnamefont {Neugebauer}},\ }\href {\doibase
  10.1103/PhysRevB.79.134106} {\bibfield  {journal} {\bibinfo  {journal}
  {Physical Review B}\ }\textbf {\bibinfo {volume} {79}},\ \bibinfo {pages}
  {134106} (\bibinfo {year} {2009})}\BibitemShut {NoStop}%
\bibitem [{\citenamefont {Zunger}\ \emph {et~al.}(1990)\citenamefont {Zunger},
  \citenamefont {Wei}, \citenamefont {Ferreira},\ and\ \citenamefont
  {Bernard}}]{Zunger1990}%
  \BibitemOpen
  \bibfield  {author} {\bibinfo {author} {\bibfnamefont {A.}~\bibnamefont
  {Zunger}}, \bibinfo {author} {\bibfnamefont {S.~H.}\ \bibnamefont {Wei}},
  \bibinfo {author} {\bibfnamefont {L.~G.}\ \bibnamefont {Ferreira}}, \ and\
  \bibinfo {author} {\bibfnamefont {J.~E.}\ \bibnamefont {Bernard}},\ }\href
  {\doibase 10.1103/PhysRevLett.65.353} {\bibfield  {journal} {\bibinfo
  {journal} {Physical Review Letters}\ }\textbf {\bibinfo {volume} {65}},\
  \bibinfo {pages} {353} (\bibinfo {year} {1990})}\BibitemShut {NoStop}%
\bibitem [{\citenamefont {Mermin}(1965)}]{Mermin1965}%
  \BibitemOpen
  \bibfield  {author} {\bibinfo {author} {\bibfnamefont {N.}~\bibnamefont
  {Mermin}},\ }\href {\doibase 10.1103/PhysRev.137.A1441} {\bibfield  {journal}
  {\bibinfo  {journal} {Physical Review}\ }\textbf {\bibinfo {volume} {137}},\
  \bibinfo {pages} {A1441} (\bibinfo {year} {1965})}\BibitemShut {NoStop}%
\bibitem [{\citenamefont {Inden}(1976)}]{FurEisenforschung1976}%
  \BibitemOpen
  \bibfield  {author} {\bibinfo {author} {\bibfnamefont {G.}~\bibnamefont
  {Inden}},\ }\href {http://books.google.se/books?id=BF4VtwAACAAJ} {\emph
  {\bibinfo {title} {{Project Meeting CALPHAD V}}}}\ (\bibinfo {year} {1976})\
  pp.\ \bibinfo {pages} {1--13}\BibitemShut {NoStop}%
\bibitem [{\citenamefont {Redlich}\ and\ \citenamefont
  {Kister}(1948)}]{Redlich1948}%
  \BibitemOpen
  \bibfield  {author} {\bibinfo {author} {\bibfnamefont {O.}~\bibnamefont
  {Redlich}}\ and\ \bibinfo {author} {\bibfnamefont {A.~T.}\ \bibnamefont
  {Kister}},\ }\href {\doibase 10.1021/ie50458a035} {\bibfield  {journal}
  {\bibinfo  {journal} {Industrial \& Engineering Chemistry}\ }\textbf
  {\bibinfo {volume} {40}},\ \bibinfo {pages} {341} (\bibinfo {year}
  {1948})}\BibitemShut {NoStop}%
\bibitem [{\citenamefont {Johnson}\ \emph {et~al.}(2012)\citenamefont
  {Johnson}, \citenamefont {Thuvander}, \citenamefont {Stiller}, \citenamefont
  {Od\'{e}n},\ and\ \citenamefont {Hultman}}]{Johnson2012}%
  \BibitemOpen
  \bibfield  {author} {\bibinfo {author} {\bibfnamefont {L.}~\bibnamefont
  {Johnson}}, \bibinfo {author} {\bibfnamefont {M.}~\bibnamefont {Thuvander}},
  \bibinfo {author} {\bibfnamefont {K.}~\bibnamefont {Stiller}}, \bibinfo
  {author} {\bibfnamefont {M.}~\bibnamefont {Od\'{e}n}}, \ and\ \bibinfo
  {author} {\bibfnamefont {L.}~\bibnamefont {Hultman}},\ }\href {\doibase
  10.1016/j.tsf.2012.02.085} {\enquote {\bibinfo {title} {{Spinodal
  decomposition of Ti0.33Al0.67N thin films studied by atom probe
  tomography}},}\ } (\bibinfo {year} {2012})\BibitemShut {NoStop}%
\bibitem [{\citenamefont {Knutsson}\ \emph {et~al.}(2013)\citenamefont
  {Knutsson}, \citenamefont {Ullbrand}, \citenamefont {Rogström},
  \citenamefont {Norrby}, \citenamefont {Johnson}, \citenamefont {Hultman},
  \citenamefont {Almer}, \citenamefont {{Johansson Jöesaar}}, \citenamefont
  {Jansson},\ and\ \citenamefont {Odén}}]{Knutsson2013}%
  \BibitemOpen
  \bibfield  {author} {\bibinfo {author} {\bibfnamefont {A.}~\bibnamefont
  {Knutsson}}, \bibinfo {author} {\bibfnamefont {J.}~\bibnamefont {Ullbrand}},
  \bibinfo {author} {\bibfnamefont {L.}~\bibnamefont {Rogström}}, \bibinfo
  {author} {\bibfnamefont {N.}~\bibnamefont {Norrby}}, \bibinfo {author}
  {\bibfnamefont {L.~J.~S.}\ \bibnamefont {Johnson}}, \bibinfo {author}
  {\bibfnamefont {L.}~\bibnamefont {Hultman}}, \bibinfo {author} {\bibfnamefont
  {J.}~\bibnamefont {Almer}}, \bibinfo {author} {\bibfnamefont {M.~P.}\
  \bibnamefont {{Johansson Jöesaar}}}, \bibinfo {author} {\bibfnamefont
  {B.}~\bibnamefont {Jansson}}, \ and\ \bibinfo {author} {\bibfnamefont
  {M.}~\bibnamefont {Odén}},\ }\href {\doibase 10.1063/1.4809573} {\bibfield
  {journal} {\bibinfo  {journal} {Journal of Applied Physics}\ }\textbf
  {\bibinfo {volume} {113}},\ \bibinfo {pages} {213518} (\bibinfo {year}
  {2013})}\BibitemShut {NoStop}%
\end{thebibliography}
%

\end{document}


\title{Anharmonicity changes the solid solubility of a random alloy at high temperatures --- supplementary information} 
\author{Nina Shulumba}
\affiliation{Department of Physics, Chemistry, and Biology (IFM), Link\"{o}ping University, SE-581 83, Link\"oping, Sweden.}
\affiliation{Functional Materials, Saarland University, Campus D3 3, D-66123 Saarbr\"{u}cken, Germany.}

\author{Olle Hellman}
\affiliation{Division of Engineering and Applied Science, California Institute of Technology, Pasadena, California 91125.}
\affiliation{Department of Physics, Chemistry, and Biology (IFM), Link\"{o}ping University, SE-581 83, Link\"oping, Sweden.}

\author{Zamaan Raza}
\affiliation{Department of Physics, Chemistry, and Biology (IFM), Link\"{o}ping University, SE-581 83, Link\"oping, Sweden.}

\author{Bj\"{o}rn Alling}
\affiliation{Department of Physics, Chemistry, and Biology (IFM), Link\"{o}ping University, SE-581 83, Link\"oping, Sweden.}
\affiliation{Max-Planck-Institut f\"{u}r Eisenforschung GmbH, D-40237 D\"{u}sseldorf, Germany}

\author{Jenifer Barrirero}
\affiliation{Functional Materials, Saarland University, Campus D3 3, D-66123 Saarbr\"{u}cken, Germany.}
\affiliation{Department of Physics, Chemistry, and Biology (IFM), Link\"{o}ping University, SE-581 83, Link\"oping, Sweden.}

\author{Frank M\"{u}cklich}
\affiliation{Functional Materials, Saarland University, Campus D3 3, D-66123 Saarbr\"{u}cken, Germany.}

\author{Igor A. Abrikosov}
\affiliation{Department of Physics, Chemistry, and Biology (IFM), Link\"{o}ping University, SE-581 83, Link\"oping, Sweden.}
\affiliation{Materials Modeling and Development Laboratory, NUST "MISIS", 119049 Moscow, Russia.}
\affiliation{LOCOMAS Laboratory, Tomsk State University, 634050, Tomsk, Russia}

\author{Magnus Od\'{e}n}
\affiliation{Department of Physics, Chemistry, and Biology (IFM), Link\"{o}ping University, SE-581 83, Link\"oping, Sweden.}

\date{\today}

\maketitle

\section{Methodological details}
In order to find the effective force constant matrix that best represents the Born-Oppenheimer potential energy surface, we minimize the difference in forces between the model system and the SQS model of a real alloy, computing the latter by means of \textit{ab initio} molecular dynamics (AIMD). We seek the linear least squares solution for the force constants $\mat{\Phi}^{\textrm{eff}}_{ij}$ that minimize the difference in forces $\Delta \vec{F}$ between AIMD and our Hamiltonian form $\set{\vec{F}_t^{\textrm{m}}}$.
\begin{equation}\label{eq:min_f}
\begin{split}
\min_{\mat{\Phi}}\Delta \vec{F} & =\frac{1}{N_t} \sum_{t=1}^{N_t}  \left| \vec{F}_t^{\textrm{MD}}-\vec{F}_t^{\textrm{H}} \right|^2 \\
& =\frac{1}{N_t} \sum_{t=1}^{N_t} \left| \vec{F}_t^{\textrm{MD}}-\mat{\Phi}^{\textrm{eff}}\vec{U}_t^{\textrm{MD}} \right|^2,
\end{split}
\end{equation}
where $\set{\vec{U}_t^{\textrm{MD}}}$ and $\set{\vec{F}_t^{\textrm{MD}}}$ are sets of displacements and forces from AIMD respectively.

TDEP potential energy surface is given by 
\begin{equation}\label{eq:harmenagain}
U_{\textrm{TDEP}}(t)= U_0(t)+\frac{1}{2} \sum_{ij} 
\vec{u}_{i}(t) \mat{\Phi}^{\textrm{eff}}_{ij}(t) \vec{u}_{j}(t),
\end{equation}
where $\vec{u}_{i,j}$ are displacement of atoms i, j respectively and $t$ is time. $U_0$ is a non-harmonic term, containing the renomalized baseline for the TDEP quasiparticles. It is determined from the criteria that the average potential energy from MD and TDEP should be equal:
\begin{equation}\label{eq:unoll_from_md}
U_0(T)=\avg{U_{\textrm{MD}}(t)-\frac{1}{2} \sum_{ij} 
\vec{u}_{i}(t) \mat{\Phi}^{\textrm{eff}}_{ij}(t) \vec{u}_{j}(t)},
\end{equation}
where the brackets indicate a configuration (time) average at temperature $T$. To extract only the non-harmonic renormalization term we take the difference between between 0 K energies and calculated $U_0$ in eq. \eqref{eq:unoll_from_md}:
\begin{equation}\label{eq:deltaunot}
\Delta U_0(T)=U_0(T)-U_{el}.
\end{equation}
This term later is added to the vibrational free energy in eq. (3) in the main text.

\section{Computational details}
\subsection{\textit{Ab initio} molecular dynamics}
Canonical ensemble \textit{ab initio} molecular dynamics simulations were performed using the projector-augmented wave (PAW) method \cite{Blochl1994} as implemented in VASP\cite{Kresse1993,Kresse1996,Kresse1996b,Kresse1999} over a range of temperatures, volumes, and concentrations. The temperature was set with a Nos\'{e} thermostat \cite{Nose1984,Hoover1985} with Fermi smearing corresponding to the simulation temperature. The plane wave energy cutoff was set to 600 eV and the Brillouin zone (BZ) was sampled at the Gamma point. The random alloy was generated using a special quasirandom structure (SQS) approach \cite{Zunger1990}. A 128 atom SQS ($4\times 4\times 4$) supercell was constructed, in which 64 atoms were metal and 64 were nitrogen. Three different SQS were generated, corresponding to different compositions of the random Ti$_{1-x}$Al$_{x}$N alloy, with B1 symmetry and $x$=$\lbrace$0.25,0.5,0.75$\rbrace$. 128 atoms SQS was generated trying to mimic the SRO parameters of perfect random alloys for the first 7 coordination shells for pair interactions with extra focus on the first 5. The small-radius 4 and 3-site clusters where taken into account, but focus was on pairs. For Al-rich compositions all configurational interactions become strong, but the short range 2-site interactions give the largest contribution to the energy. We assume that the vibrational free energy shows a qualitatively similar dependence with respect to energy on 2-site correlation functions as it does with 3- and 4-site functions.

A subset of uncorrelated samples from the AIMD simulations was selected and upsampled to high accuracy with a $3\times 3\times 3$ k-point mesh for the BZ integration. The effective force constants where found to be smooth and easily interpolated across the whole concentration interval, giving the phonon DOS as a continuous function of concentration, volume and temperature. 

\subsection{Thermodynamic integration}
Thermodynamic integration was used to validate our choice of Hamiltonian. A Langevin thermostat was used to control the temperature and break the mode-locking. The numerical integration over coupling parameter $\lambda$ was carried out over 5 discrete steps, and the MD simulations were run for $\sim$100 000 timesteps with k-point sampling at the Gamma-point only. A set of uncorrelated snapshots was selected from AIMD and upsampled to high accuracy with a $3\times 3\times 3$ k-point mesh to ensure convergence for $\Delta F$.

\subsection{Benchmark test}
\begin{figure}
\includegraphics[width=\linewidth]{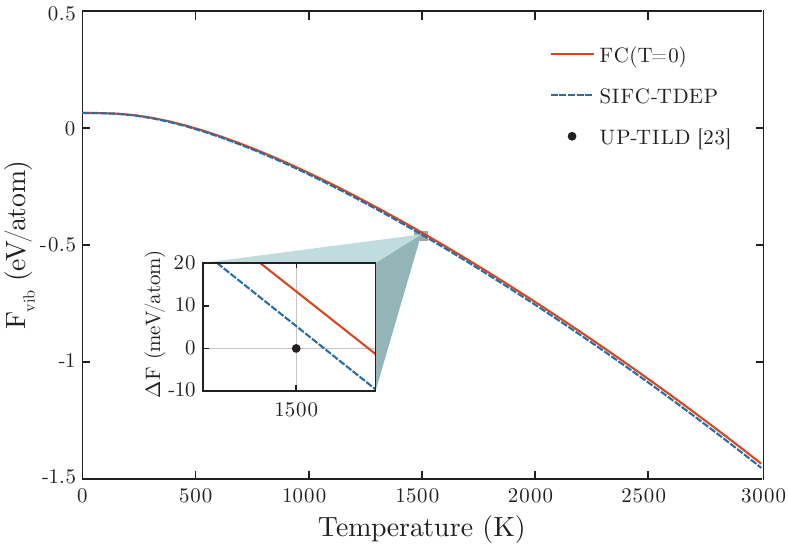}
\caption{\label{fig:fvib} (Color online) Phonon free energy as a function of temperature at constant volume. Harmonic free energy calculated with force constants extracted at 0K using small displacement method (red solid line). Harmonic free energy calculated using SIFC-TDEP with the force constants and $\Delta U_0$ at the same volume (1500 K) is shown by blue dashed line. Free energy calculated by upsampled thermodynamic integration using Langevin dynamics is shown as the black point. We include the insert showing the difference between mentioned above free energies at 1500 K taking the thermodynamic integration point as a reference.}
\end{figure}

We perform a series of additional calculations to check the reliability of our method. Using the small displacement method, the harmonic vibrational Gibbs free energy was calculated for the full SQS, treating all atoms as non-equivalent and thus including local environment effects in the force constants. This was compared to the SIFC-TDEP vibrational Gibbs free energy (Fig. 1 in the main text). Harmonic phonon free energies are compared at a specific concentration (Ti$_{0.5}$Al$_{0.5}$N) and volume/atom (9.1260 \AA$^3$). For the small displacement method, force constants are taken at zero temperature, while ``SIFC-TDEP'' denotes a free energy calculated using SIFC-TDEP force constants. At high temperature, the harmonic approximation is no longer a good benchmark, even at constant volume. In this case, we benchmarked using UP-TILD, which is formally exact in that region. \figref{fig:fvib} shows the comparison of the phonon free energies at the same volume taken at 1500 K. The free energy calculated using temperature-dependent force constants and the anharmonic correction ($\Delta U_0$) lies closer to the thermodynamic integration point. Omitting anharmonic effects leads to the underestimation of the phonon contribution (\figref{fig:fvib}).

\section{Phase diagram reconstruction}
G$_{mix}(\vec{x})$ is defined as a function of the vector $\vec{x}$ in composition space. In the general case at a given pressure and temperature, a mixture of $N_c$ atomic species, or components, may split into $N_p$ distinct phases. We define a $N_c \times N_p-1$ grid of compositions $\vec{x}$ with elements $x_{ik}$, where element $x_{ik}$ is an atomic fraction of component $i$ in phase $k$. The solution must satisfy two constraints \cite{Gibbs1871} : the component material balance constraint and the positive value of the atomic fraction of components in the system:
\begin{equation}
\begin{aligned}
&\sum_{i=1}^{N_c}\sum_{k=1}^{N_p-1} x_{ik} = 1,\\
& 0 \leq x_{ik} \leq x_{i}.
\end{aligned}
\end{equation}

The Gibbs free energy can be expressed as discrete function
\begin{equation}
G(\vec{x})=\sum\limits_{i=1}^{N_c} \sum\limits_{k=1}^{N_p-1} x_{ik}G_{ik}. 
\end{equation}

The mass balance constraint can be rewritten in matrix form $\vec{A}\vec{x}=\vec{B}$, or
\begin{equation}\label{eq:constr}
\sum\limits_{i=1}^{N_c}\sum\limits_{k=1}^{N_p-1} a_{ei} x_{ik}=B,
\end{equation}
where $e$ is an index of number of possible combinations of fractions of component $i$ and a$_{ei}$ are the matrix elements. B is a global concentration of component $i$ in the initial solid solution phase.

We consider the Gibbs free energy of mixing. There are two phases or distinct regions (k=2); one is a solid solution of Ti$_{1-x}$Al$_{x}$N and the other is the miscibility region.
\begin{equation}
G_{mix}(\vec{x})=\sum\limits_{i=1}^{N_c} G^{mix}_{i}. 
\end{equation}

To recover the phase diagram, we minimise $G_\mathrm{mix}$ or $G$ subject to the constraint \eqref{eq:constr}.

This can be done via, for example, the method of constrained least squares \cite{antoniou2007practical}. In order to construct the phase diagram, we repeat this procedure over a range of temperatures in the regime of interest.

This is a convenient methodology since no assumptions about reactions are necessary, and the constraints are easy to identify. There is also no need to define $G$ across the entire concentration interval if we have a many phase system, which removes complications such as the presence of dynamically unstable phases. Furthermore the proposed technique removes the need to numerically calculate derivatives of the Gibbs free energy needed in state-of-the-art methods for constructing phase diagrams, a procedure that is notoriously sensitive to numerical noise. Of course in our case, we still determine the spinodal from the condition ($\frac{\partial^2 G_{mix}}{\partial x^2}|_{T} \leq 0$).

\section{Experimental details}

Atom probe tomography (APT) samples were prepared in a dual-beam focused ion beam/scanning electron microscopy workstation (FIB/SEM) (Helios NanoLab 600$^\textrm{TM}$, FEI Company, USA) by the standard lift-out technique \cite{Thompson2007}. Laser Pulsed APT was carried out with a LEAP$^\textrm{TM}$ 3000X HR (CAMECA) at a repetition rate of 200 kHz, a specimen temperature of about 60 K, a pressure lower than 1 $\times$ 10$^{-10}$ Torr (1.33 $\times$ 10$^{-8}$ Pa), and a laser pulse energy of 0.5 nJ. The evaporation rate of the specimen was 5 atoms per 1000 pulses. The positions and the time-of-flight of the ions coming into the detector are used to generate a three-dimensional reconstruction with IVAS$^\textrm{TM}$3.6.6 software (CAMECA). The accuracy of atomic-scale of APT and concentration are discussed in literature \cite{Seidman2007,Allen2008,Kim2014,Biswas2014}. Previous results demonstrate that it is now possible to obtain highly quantitative information from APT.

%